\documentclass[journal]{IEEEtran}
\usepackage{cite}

\ifCLASSINFOpdf
\else
\fi
\usepackage{graphicx}

\begin{document}
%
\title{Task Oriented Video Coding: A Survey}

\author{Daniel Wood, Member, IEEE}
        
\maketitle

\begin{abstract}
Video coding technology has been continuously improved for higher compression ratio with higher resolution. However, the state-of-the-art video coding standards, such as H.265/HEVC and Versatile Video Coding, are still designed with the assumption the compressed video will be watched by humans. With the tremendous advance and maturation of deep neural networks in solving computer vision tasks, more and more videos are directly analyzed by deep neural networks without humans' involvement. Such a conventional design for video coding standard is not optimal when the compressed video is used by computer vision applications. While the human visual system is consistently sensitive to the content with high contrast, the impact of pixels on computer vision algorithms is driven by specific computer vision tasks. In this paper, we explore and summarize recent progress on computer vision task oriented video coding and emerging video coding standard, Video Coding for Machines.
\end{abstract}


\IEEEpeerreviewmaketitle

\section{Introduction}

\IEEEPARstart{V}{ideo} coding technologies and standards are based on a set of principles that reduce the visual information redundancy in digital video for human eyes. The block-based hybrid approach (intra-picture and inter-picture prediction and 2D transform coding) is employed in all modern video compression standards \cite{roese1975combined} include H.261 \cite{liou1991overview}, H.262/MPEG-2 Video \cite{union1994generic}, H.263 \cite{rijkse1996h}, MPEG-1 \cite{yoshioka1999information}, MPEG-4 Visual \cite{ebrahimi2000mpeg}, and H.264/AVC \cite{sullivan2004h}. Common video codecs can deliver a certain level quality for humans in a low bit rate, which is achieved by rate-distortion optimization \cite{sullivan1998rate}. In terms of video streaming, the quality of experience is proposed to measure and adjust the perceived video quality by end user \cite{sadat2019qoe} \cite{kong2017impact}. As higher resolution such as 4K UHD and 8K UHD are more and more popular in this big data era, more challenges and higher compression rate are expected for video coding technology. Thanks to continuous improvement of video coding technology, higher compression ratio compared with raw video is achieved. However, the state-of-the-art video coding standards, such as H.265/HEVC \cite{sullivan2012overview} and Versatile Video Coding (VCC) \cite{bross2021overview}, are still designed with the assumption the compressed video will be watched by humans.

On the other hand, deep neural networks have advanced tremendously and matured over the past few years in commercial adoption on both powerful GPU servers and computation constrained mobile devices with CPUs. Representative algorithms include one-stage fast object detection framework such as You Only Look Once (YOLO) \cite{redmon2016you} and Single Shot Detector (SSD) \cite{liu2016ssd} and two-stage complicated object detection framework such as Fast Region-based Convolutional Neural Network (Fast R-CNN) \cite{girshick2015fast} and Faster R-CNN \cite{ren2015faster}. These popular algorithms can achieve either fast detection or accuracy detection, even object segmentation by Mask R-CNN \cite{he2017mask} and object tracking \cite{wu2013online}.

The computer vision tasks are implemented either in the central server or in the edge device \cite{kong2019blind} \cite{redondi2016compress} . In the Compress-Then-Analyze (CTA) paradigm, images acquired from camera are compressed and sent to a central server for further analysis. Conversely, in the Analyze-Then-Communicate (ATC) approach, edge devices perform object detection and tracking, semantic feature extraction, and communicate semantic information to the network \cite{kong2019no}. Such ATC approach is also called Edge AI. In this paper, we will focus on the first scenario namely Compress-Then-Analyze.

Due to massive amount and length of video, more and more videos are not observed by humans anymore, but directly analyzed by deep neural networks on the local machines or on the cloud. Such a conventional design for video coding standard is not optimal when the compressed video is used by computer vision applications. Characteristics of the human visual system (HVS) such as the visual attention and the contrast sensitivity mechanisms are considered in perceptual image quality assessment \cite{wang2004image, ma2021optimal, ma2020adversarial}. Due to the foveation feature of the HVS, at an instance, only a local area in the image can be perceived with high resolution at typical viewing distances, and the HVS is sensitive to the relative rather than the absolute luminance change \cite{kong2021image}. Unfortunately, a simple combination of domain adaptation and semi-supervised learning methods often fail to address such two objects because of training data bias towards labeled samples \cite{wang2019inductive, ma2020optimal, qin2021semi, ma2022first}. An adaptive structure learning method to regularize the cooperation of semi-supervised learning \cite{gao2021finite, ma2017improved} and domain adaptation \cite{qin2021contradictory, ma2016new} might be an efficient approach. While the HVS is consistently sensitive to the content with high contrast, the impact of pixels on computer vision algorithms is task driven.

It is worth mentioning that there is a terminology called "machine vision". There is some confusion between machine vision and computer vision. Actually, machine vision came first \cite{steger2018machine}. This engineering-based system uses existing technologies to mechanically ‘see’ steps along a production line. It helps manufacturers detect flaws in their products before they are packaged, or food distribution companies ensure their foods are correctly labelled, for instance. Such applications in industrial manufacturing are call anomaly detection or defect detection.

In this paper, we will explore and summarize recent progress on computer vision task oriented video coding. In Section \ref{VCforMachines}, the relevant video coding standard, namely Video Coding for Machines (VCM), is introduced. In Section \ref{VCforTask}, the block-based hybrid video coding schemes designed for specific computer vision tasks are introduced and discussed. We conclude the trend in \ref{end}.

\section{Video Coding for Machines}
\label{VCforMachines}

\begin{figure*}
    \centering
    \includegraphics[width=\textwidth]{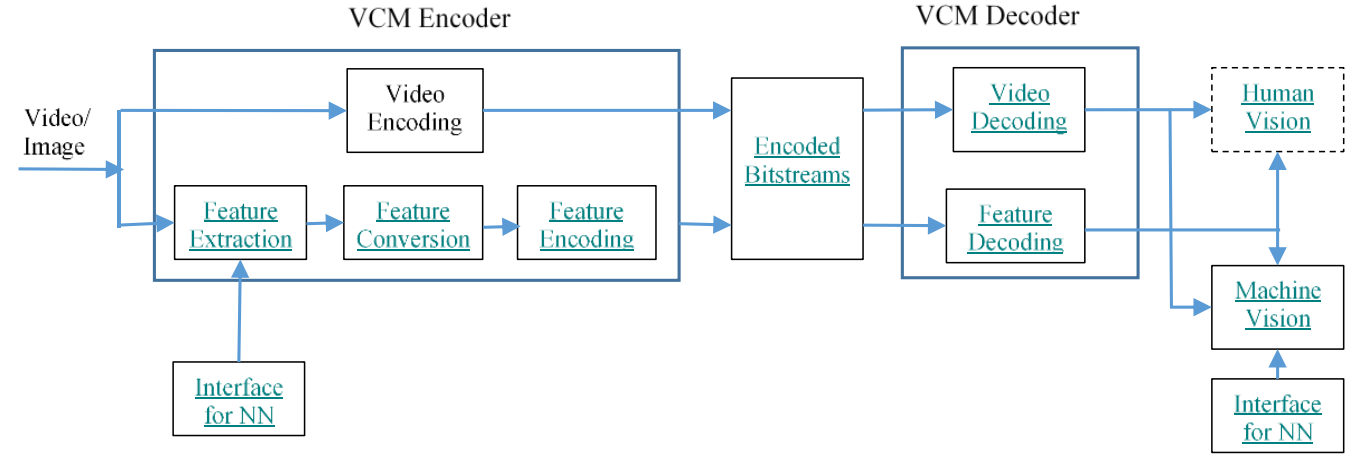}
    \caption{The potential VCM architecture \cite{gao2021recent}}
\label{VCM}
\end{figure*}

Video coding, which targets to compress and reconstruct the whole frame, and feature compression, which only preserves and transmits the most critical information, stand at
two ends of the scale. That is, one is with compactness and efficiency to serve for computer vision, and the other is with full fidelity, bowing to human perception. The recent endeavors in imminent trends of video compression, e.g. deep learning based coding tools and end-to-end image/video coding, and MPEG-7 compact feature descriptor standards, i.e. Compact Descriptors for Visual Search \cite{duan2015overview} (CDVS) and Compact Descriptors for Video Analysis \cite{duan2018compact} (CDVA), promote the sustainable and fast development in their own directions, respectively. In CDVS, hand-crafted local and global descriptors are designed to represent the visual characteristics of images. CDVS has made remarkable improvements over a large-scale benchmark in image retrieval performance with 6 compact feature descriptors. Multiple key techniques, e.g. light-weighted interest point selection \cite{gao2018data} \cite{francini2013selection}, low-degree polynomial detector \cite{garbo2015accurate, ma2017new, li2016new, ma2017stability}, location histogram coding \cite{tsai2009location} \cite{tsai2012improved}, and scalable compressed fisher vector \cite{wu2018codebook} \cite{lin2014rate}, have been developed by collaborative experiments within a evaluation framework. In CDVA, the deep learning features are adopted to further boost the video analysis performance. In the experiments \cite{lin2017hnip}, comparable performances of the deep learning features with different off-the-shelf deep neural networks models (such as AlexNet \cite{krizhevsky2012imagenet}, VGG-16 \cite{simonyan2014very}, ResNet \cite{he2016deep}) are reported. Due to the hardware friendly merits and the competitive performance, VGG-16 model is adopted by CDVA as the normative backbone network to derive deep feature representations. Thanks to booming AI technology, authors in the \cite{duan2020video} carry out exploration in the new area, Video Coding for Machines (VCM), arising from the emerging MPEG standardization efforts. Towards collaborative compression and intelligent analytics, VCM attempts to bridge the gap between feature coding for machine and video coding for human vision. Aligning with the rising Analyze then Compress instance Digital Retina, the definition, formulation, and paradigm of VCM are given first. Meanwhile, we systematically review state-of-the-art techniques in video compression and feature compression from the unique perspective of MPEG standardization, which provides the academic and industrial evidence to realize the collaborative compression of video and feature streams in a broad range of AI applications. Finally, authors come up with potential VCM solutions, and the preliminary results have demonstrated the performance and efficiency gains.

In paper \cite{hu2020towards}, authors come up with a novel image coding framework by leveraging both the compressive and the generative models, to support computer vision and human perception tasks jointly. Given an input image, the feature analysis is first applied, and then the generative model is employed to perform image reconstruction with features and additional reference pixels, in which compact edge maps are extracted in this work to connect both kinds of vision in a scalable way. The compact edge map serves as the basic layer for computer vision tasks, and the reference pixels act as a sort of enhanced layer to guarantee signal fidelity for human vision. By introducing advanced generative models, authors train a flexible network to reconstruct images from compact feature representations and the reference pixels. Experimental results demonstrate the superiority of our framework in both human visual quality and facial landmark detection, which provide useful evidence on the emerging standardization efforts on MPEG VCM.

With the rapid advances of deep feature representation and visual data compression in mind, in \cite{yang2021video}, authors summarize VCM methodology and philosophy based on existing academia and industrial efforts. The development of VCM follows a general rate-distortion optimization, and the categorization of key modules or techniques is established including features assisted coding, scalable coding, intermediate feature compression/optimization, and computer vision targeted codec, from broader perspectives of vision tasks, analytics resources, etc. From previous works, it is demonstrated that, although existing works attempt to reveal the nature of scalable representation in bits when dealing with machine and human vision tasks, there remains a rare study in the generality of low bit rate representation, and accordingly how to support a variety of visual analytic tasks. Therefore, authors investigate a novel visual information compression for the analytics taxonomy problem to strengthen the capability of compact visual representations extracted from multiple tasks for visual analytics. A new perspective of task relationships versus compression is revisited. By keeping in mind the transferability among different computer vision tasks (e.g. high-level semantic and mid-level geometry-related), authors aim to support multiple tasks jointly at low bit rates. In particular, to narrow the dimensionality gap between neural network generated features extracted from pixels and a variety of computer vision features/labels (e.g. scene class, segmentation labels), a codebook hyperprior is designed to compress the neural network-generated features. As demonstrated in their experiments, the new hyperprior model is expected to improve feature compression efficiency by estimating the signal entropy more accurately, which enables further investigation of the granularity of abstracting compact features among different tasks.

Efficient VCM has become an important topic in academia and industry. Recent standard development activities on VCM are introduced in \cite{gao2021recent}. In July 2019, the international standardization organization, i.e., MPEG, created an Ad-Hoc group named VCM to study the requirements for potential standardization work. The MPEG VCM activity aims to standardize a bitstream format generated by compressing a previously extracted feature stream or video stream. Fig. \ref{VCM} shows an example of potential VCM architecture. The VCM codec could be a video codec or a feature codec, or both. In the case of feature codec, the VCM feature encoding consists of feature extraction, feature conversion/packing, and feature coding. There may be an interface to an external Neural Network Compression and Representation for the feature extraction and the task-specific networks. The MPEG VCM group has identified six classes of use cases \cite{zhang2021use} including Intelligent Transportation, Smart City, and Intelligent Contents. The MPEG VCM group has established an evaluation framework that includes computer vision tasks \cite{rafie2021evaluation}. Typical tasks include object detection, object segmentation, object tracking, action recognition, and pose estimation.

The work \cite{mackowiak2022video} deals with Video Coding for Machines that is a new paradigm in video coding related to consumption of decoded video by humans and machines. For such tasks, joint transmission of compressed video and features is considered. In this paper, we focus our considerations of features on SIFT keypoints. They can be extracted from the decoded video with losses in number of key points and their parameters as compared to the SIFT key points extracted from the original video. Such losses are studied for HEVC and VVC as functions of the quantization parameter and the bitrate. Authors propose to transmit the residual feature data together with the compressed video. Therefore, even for strongly compressed video, the transmission of whole all SIFT keypoint information is avoided. 

Traditional media coding schemes typically encode image/video into a semantic-unknown binary stream, which fails to directly support downstream intelligent tasks at the bitstream level. Semantically Structured Image Coding (SSIC) framework \cite{sun2020semantic} makes the first attempt to enable decoding-free or partial-decoding image intelligent task analysis via a Semantically Structured Bitstream (SSB). However, the SSIC only considers image coding and its generated SSB only contains the static object information. In this work \cite{jin2022semantically}, authors extend the idea of semantically structured coding from video coding perspective and propose an advanced Semantically Structured Video Coding (SSVC) framework to support heterogeneous intelligent applications. Video signals contain more rich dynamic motion information and exist more redundancy due to the similarity between adjacent frames. Thus, we present a reformulation of semantically structured bitstream (SSB) in SSVC which contains both of static object characteristics and dynamic motion clues. Specifically, we introduce optical flow to encode continuous motion information and reduce cross-frame redundancy via a predictive coding architecture, then the optical flow and residual information are reorganized into SSB, which enables the proposed SSVC could better adaptively support video-based downstream intelligent applications. Extensive experiments demonstrate that the proposed SSVC framework could directly support multiple intelligent tasks just depending on a partially decoded bitstream. This avoids the full bitstream decompression and thus significantly saves bitrate/bandwidth consuming for intelligent analytics. We verify this point on the tasks of image object detection, pose estimation, video action recognition, video object segmentation, etc.

Since standard image compression techniques such as JPEG or JPEG2000 (J2K) are often designed to maximize image quality as measured by conventional quality metrics such as mean-squared error (MSE) or Peak Signal to Noise Ratio (PSNR). This mismatch between image quality metrics and ask performance motivates our investigation of image compression using a task-specific metric designed for the designated tasks. Given the selected target detection task, authors in \cite{pu2014image} first propose a metric based on conditional class entropy. The proposed metric is then incorporated into a J2K encoder to create compressed codestreams that are fully compliant with the J2K standard. Experimental results illustrate that the decompressed images obtained using the proposed encoder greatly improve performance in detection/classification tasks over images encoded using a conventional J2K encoder.

To reduce the data storage and transfer overhead in smart resource-limited Internet-of-Thing (IoT) systems, effective data compression is a "must-have" feature before transferring real-time produced dataset for training or classification. Authors in \cite{liu2018deepn} develop an image compression framework tailored for DNN applications to embrace the nature of deep cascaded information process mechanism of DNN architecture. Extensive experiments, based on "ImageNet" dataset with various state-of-the-art DNNs, show that "DeepN-JPEG" can achieve around 3.5× higher compression rate over the popular JPEG solution while maintaining the same accuracy level for image recognition, demonstrating its great potential of storage and power efficiency in DNN-based smart IoT system design.

\cite{chen2019learning} proposes a Learning based Facial Image Compression (LFIC) framework with a novel Regionally Adaptive Pooling module whose parameters can be automatically optimized according to gradient feedback from an integrated hybrid semantic fidelity metric, including a successfully exploration to apply Generative Adversarial Network (GAN) as metric directly in image compression scheme. The experimental results verify the framework’s efficiency by demonstrating performance improvement of 71.41\%, 48.28\% and 52.67\% Bit Rate Saving separately over JPEG2000, WebP and neural network-based codecs under the same face verification accuracy distortion metric. We also evaluate LFIC’s superior performance gain compared with latest specific facial image codecs. Visual experiments also show some interesting insight on how LFIC can automatically capture the information in critical areas based on semantic distortion metrics for optimized compression, which is quite different from the heuristic way of optimization in traditional image compression algorithms.

In \cite{xia2020emerging}, authors study a new problem arising from the emerging MPEG standardization effort VCM, which aims to bridge the gap between visual feature compression and classical video coding. VCM is committed to address the requirement of compact signal representation for both machine and human vision in a more or less scalable way. To this end, authors make endeavors in leveraging the strength of predictive and generative models to support advanced compression techniques for both machine and human vision tasks simultaneously, in which visual features serve as a bridge to connect signal-level and task-level compact representations in a scalable manner. Specifically, we employ a conditional deep generation network to reconstruct video frames with the guidance of learned motion pattern. By learning to extract sparse motion pattern via a predictive model, the network elegantly leverages the feature representation to generate the appearance of to-be-coded frames via a generative model, relying on the appearance of the coded key frames. Meanwhile, the sparse motion pattern is compact and highly effective for high-level vision tasks, e.g. action recognition. Experimental results demonstrate that our method yields much better reconstruction quality compared with the traditional video codecs (0.0063 gain in SSIM), as well as state-of-the-art action recognition performance over highly compressed videos (9.4\% gain in recognition accuracy), which showcases a promising paradigm of coding signal for both human and computer vision.

The ubiquitous camera networks in the city brain system grow at a rapid pace, creating massive amounts of images and videos at a range of spatial-temporal scales and thereby forming the “biggest” big data. However, the sensing system often lags behind the construction of the fast-growing city brain system, in the sense that such exponentially growing data far exceed today’s sensing capabilities. Therefore, critical issues arise regarding how to better leverage the existing city brain system and significantly improve the city-scale performance in intelligent applications. To tackle the unprecedented challenges, authors in \cite{gao2021digital} articulate a vision towards a novel visual computing framework, termed as digital retina , which aligns high-efficiency sensing models with the emerging VCM paradigm. In particular, digital retina may consist of video coding, feature coding, model coding, as well as their joint optimization. The digital retina is biologically-inspired, rooted on the widely accepted view that the retina encodes the visual information for human perception, and extracts features by the brain downstream areas to disentangle the visual objects. Within the digital retina framework, three streams, i.e., video stream, feature stream, and model stream, work collaboratively over the end-edge-cloud platform. In particular, the compressed video stream serves for human vision, the compact feature stream targets for computer vision, and the model stream incrementally updates deep learning models to improve the performance of human/computer vision tasks. Authors have developed a prototype to demonstrate the technical advantages of digital retina, and extensive experiments have been conducted to validate that it is able to effectively support the video big data analysis and retrieval in the intelligent city system.

\section{Video Coding Scheme for Specific Computer Vision Tasks}
\label{VCforTask}

The vigorous developments of the Internet of Things make it possible to extend its computing and storage capabilities to computing tasks in the aerial system with the collaboration of cloud and edge, especially for artificial intelligence (AI) tasks based on deep learning. Collecting a large amount of image/video data, unmanned aerial vehicles (UAVs) can only hand over intelligent analysis tasks to the back-end mobile edge computing server due to their limited storage and computing capabilities. How to efficiently transmit the most correlated information for the AI model is a challenging topic. Inspired by task-oriented communication in recent years, authors in \cite{kang2022task} propose a new aerial image transmission paradigm for the scene classification task . A lightweight model is developed on the front-end UAV for semantic block transmission with the perception of images and channel states. To achieve the trade-off between transmission latency and classification accuracy, deep reinforcement learning is applied to explore the semantic blocks which have the greatest contribution to the back-end classifier under various channel states. The proposed method can significantly improve classification accuracy by more than 4\% under the same conditions, compared to other semantic saliency learning methods.

Human oriented video coding strategies may not be optimal when the compressed signal is further processed by algorithms, as the encoder is unaware of the task specific information. For example, because of the different categories of objects used to train detection algorithms, the influence of the same image content on those detectors also varies \cite{kong2016new}. Authors in \cite{cai2021novel}, taking object detection as an example, propose a novel video coding strategy for computer vision. By protecting the information according to its importance for an object detector rather than for the human visual system, the proposed method has the potential to achieve a better object detection performance with the same bandwidth. The main contributions of the paper can be summarized: 1) the modeling of the relationship between object detection accuracy and bit rate; 2) a back propagation based method to analyze the influence of each pixel on the detection of target objects; 3) an object detection oriented bit allocation and codec control parameter determination scheme; 4) an evaluation metric to compare the impact of video coding strategies on a given object detector over a predefined range of bit rate. The proposed algorithm can better preserve the video content vital for object detection than state-of-the-art video coding schemes. 

Common state-of-the-art video codecs are optimized to deliver a low bitrate by providing a certain quality for the final human observer, which is achieved by rate-distortion optimization (RDO). With the steady improvement of neural networks solving computer vision tasks, more and more multimedia data is directly analyzed by neural networks. Authors in \cite{fischer2020video} propose a standard-compliant Feature-based RDO (FRDO) that is designed to increase the coding performance, when the decoded frame is analyzed by a neural network in a video coding for machine scenario. To that extent, we replace the pixel-based distortion metrics in conventional RDO of VTM-8.0 with distortion metrics calculated in the feature space created by the first layers of a neural network. Throughout several tests with the segmentation network Mask R-CNN and single images from the Cityscapes dataset, the proposed FRDO and its Hybrid version FRDO (HFRDO) with different distortion measures in the feature space are compared with the conventional RDO. With HFRDO, up to 5.49\% bitrate can be saved compared to the VTM-8.0 implementation in terms of Bjontegaard Delta Rate and using the weighted average precision as quality metric. Additionally, allowing the encoder to vary the quantization parameter results in coding gains for the proposed HFRDO of up 9.95\% compared to conventional VTM. 

In \cite{zhao2017intelligent}, authors aim at proposing an efficient, standard-compatible and simultaneously analysis-friendly coding framework for intelligent surveillance videos. The foreground objects are first extracted from the video sequence by accurate background modeling \cite{kong2016temporal}. Subsequently, the foregrounds can be constructed as a sequence and compressed in higher quality while very few background pictures are required to signal in lower quality. At the decoder side, the foreground objects can be directly used for efficient analysis tasks and the surveillance videos can be also reconstructed by synthesizing background and foreground frames. The effectiveness and potential of the proposed framework have been demonstrated in the pedestrian detection application, where the coding bits can be greatly saved with the detection accuracy being well maintained. 

Although more and more videos are delivered for video analysis (e.g., object detection/tracking and action recognition), most existing wireless video transmission schemes are developed to optimize human perception quality and are suboptimal for video analysis. In mobile surveillance networks, a cloud server collects videos from multiple moving cameras and detects suspicious persons in all camera views. Camera mobility in smartphones or dash cameras implies that video is to be uploaded through bandwidth-limited and error-prone wireless networks, which may cause quality degradation of the decoded videos and jeopardize the performance of video analyses. In \cite{chen2016quality}, authors propose an effective rate-allocation scheme for multiple moving cameras in order to improve human detection (content) performance. Therefore, the optimization criterion of the proposed rate-allocation scheme is driven by quality of content (QoC). Both video source coding and application layer forward error correction coding rates are jointly optimized. Moreover, the proposed rate-allocation problem is formulated as a convex optimization problem and can be efficiently solved by standard solvers. Many simulations using HEVC standard compression of video sequences and the deformable part model object detector are carried, and results demonstrate the effectiveness and favorable performance of the proposed QoC-driven scheme under different pedestrian densities and wireless conditions. 

Large-scale, high-quality image and video data sets significantly empower learning-based computer vision models, especially deep-learning models. However, images and videos are usually compressed before being analyzed in practical situations where transmission or storage is limited, leading to a noticeable performance loss of vision models. Author in \cite{zhang2021just} broadly investigate the impact on the performance of computer vision from image and video coding. Based on the investigation, we propose Just Recognizable Distortion (JRD) to present the maximum distortion caused by data compression that will reduce the computer vision model performance to an unacceptable level. A large-scale JRD-annotated data set containing over 340,000 images is built for various computer vision tasks, where the factors for different JRDs are studied. Furthermore, an ensemble-learning-based framework is established to predict the JRDs for diverse vision tasks under few- and non-reference conditions, which consists of multiple binary classifiers to improve the prediction accuracy. Experiments prove the effectiveness of the proposed JRD-guided image and video coding to significantly improve compression and computer vision performance. Applying predicted JRD is able to achieve remarkably better computer vision task accuracy and save a large number of bits. 

Saliency-driven image and video coding for humans has gained importance in the recent past. In \cite{fischer2021saliency} authors propose a saliency-driven coding framework for the video coding for machines task using the latest video coding standard Versatile Video Coding (VVC). To determine the salient regions before encoding, we employ the real-time-capable object detector YOLO in combination with a novel decision criterion. To measure the coding quality for a machine, the state-of-the-art object segmentation network Mask R-CNN was applied to the decoded frame. From extensive simulations we find that, compared to the reference VVC with a constant quality, up to 29\% of bitrate can be saved with the same detection accuracy at the decoder side by applying the proposed saliency-driven framework.

Thanks to the advances in computer vision systems more and more videos are going to be watched by algorithms, e.g. implementing video surveillance systems or performing automatic video tagging. This work \cite{galteri2018video} describes an adaptive video coding approach for computer vision-based systems. Through preserving the elements of the scene that are more likely to contain meaningful content improves their detection performance. The approach is based on computation of saliency maps exploiting a fast objectness measure. The computational efficiency of this approach makes it usable in a real-time video coding pipeline. Experiments show that the technique outperforms standard H.265 in speed and coding efficiency, and can be applied to different types of video domains, from surveillance to web videos. 

Video coding for satisfying humans as the final user is a widely investigated field of studies for visual content \cite{kong2017object}. But, are the assumptions and optimizations also valid when the compressed video stream is analyzed by a machine? To answer this question, authors in \cite{fischer2020intra} compare the performance of two state-of-the-art neural detection networks when being fed with deteriorated input images coded with HEVC and VVC in an autonomous driving scenario using intra coding. Additionally, the impact of the three VVC in-loop filters when coding images for a neural network is examined. The results are compared using the mean average precision metric to evaluate the object detection performance for the compressed inputs. Throughout these tests, authors found that the Bjontegaard Delta Rate savings with respect to PSNR of 22.2\% using VVC instead of HEVC cannot be reached when coding for object detection networks with only 13.6\% in the best case.

It is still attractive and challenging to implement task-driven semantic coding with the traditional hybrid coding framework, which should still be widely used in practical industry for a long time. To solve this challenge, authors in \cite{li2021task} design semantic maps for different tasks to extract the pixelwise semantic fidelity for videos/images. Instead of directly integrating the semantic fidelity metric into traditional hybrid coding framework, we implement task-driven semantic coding by implementing semantic bit allocation based on reinforcement learning (RL). We formulate the semantic bit allocation problem as a Markov decision process (MDP) and utilize one RL agent to automatically determine the quantization parameters (QPs) for different coding units (CUs) according to the task-driven semantic fidelity metric. Extensive experiments on different tasks, such as classification, detection and segmentation, have demonstrated the superior performance of our approach by achieving an average bitrate saving of 34.39\% to 52.62\% over H.265/HEVC anchor under equivalent task-related semantic fidelity.

In many distributed wireless surveillance applications, compressed videos are used for performing automatic video analysis tasks. The accuracy of object detection, which is essential for various video analysis tasks, can be reduced due to video quality degradation caused by lossy compression \cite{kong2019modeling}. The work \cite{kong2018efficient} introduces a video encoding framework with the objective of boosting the accuracy of object detection for wireless surveillance applications. The proposed video encoding framework is based on systematic investigation of the effects of lossy compression on object detection. It has been found that current standardized video encoding schemes cause temporal domain fluctuation for encoded blocks in stable background areas and spatial texture degradation for encoded blocks in dynamic foreground areas of a raw video, both of which degrade the accuracy of object detection. Two measures, the sum-of-absolute frame difference (SFD) and the degradation of texture in 2D transform domain (TXD), are introduced to depict the temporal domain fluctuation and the spatial texture degradation in an encoded video, respectively. The proposed encoding framework is designed to suppress unnecessary temporal fluctuation in stable background areas and preserve spatial texture in dynamic foreground areas based on the two measures, and it introduces new mode decision strategies for both intra and interframes to improve the accuracy of object detection while maintaining an acceptable rate distortion performance. Experimental results show that, compared with traditional encoding schemes, the proposed scheme improves the performance of object detection and results in lower bit rates and significantly reduced complexity with comparable quality in terms of PSNR and SSIM.

In \cite{choi2022scalable}, authors propose a scalable video coding framework that supports computer vision (specifically, object detection) through its base layer bitstream and human vision via its enhancement layer bitstream. The proposed framework includes components from both conventional and deep neural network-based video coding. The results show that on object detection, the proposed framework achieves 13-19\% bit savings compared to state-of-the-art video codecs, while remaining competitive in terms of MS-SSIM on the human vision task.

Video coding algorithms encode and decode an entire video frame while feature coding techniques only preserve and communicate the most critical information needed for a given application. This is because video coding targets human perception, while feature coding aims for computer vision tasks. Recently, attempts are being made to bridge the gap between these two domains. In work \cite{ahmmed2021human}, authors propose a video coding framework by leveraging on to the commonality that exists between human vision and computer vision applications using cuboids. This is because cuboids, estimated rectangular regions over a video frame, are computationally efficient, has a compact representation and object centric. Such properties are already shown to add value to traditional video coding systems. Herein cuboidal feature descriptors are extracted from the current frame and then employed for accomplishing a computer vision task in the form of object detection. Experimental results show that a trained classifier yields superior average precision when equipped with cuboidal features oriented representation of the current test frame. Additionally, this representation costs 7\% less in bit rate if the captured frames are need be communicated to a receiver.

The characteristics of human perception are not fully exploited by the conventional quantization control methods. Just Noticeable Distortion (JND)-methods facilitate this by finding perceivable distortion levels. However, these methods fail to use human attention to further reduce bitrate. To address this issue, this work \cite{nami2020juniper} proposes a Just Noticeable distortion-based Perceptual (JUNIPER) framework for video coding, which combines saliency and JND concepts. Firstly, a JND predictor is trained using Support Vector Machines (SVM) and various quality metrics. Secondly, an efficient saliency model is used to measure the visual importance of each Coding Tree Unit (CTU) in HEVC. Finally, a quantization control algorithm is proposed that jointly considers the visual importance, and the perceivable quality levels, to assign a quantization parameter (QP) to each CTU. Experimental results demonstrate that JUNIPER saves 29.57\% bitrate on average, with similar visual quality, compared with baseline HEVC coding.

In \cite{ma2018joint}, authors provide a systematical overview and analysis on the joint feature and texture representation framework, which aims to smartly and coherently represent the visual information with the front-end intelligence in the scenario of video big data applications. In particular, authors first demonstrate the advantages of the joint compression framework in terms of both reconstruction quality and analysis accuracy. Subsequently, the interactions between visual feature and texture in the compression process are further illustrated. Finally, the future joint coding scheme by incorporating the deep learning features is envisioned, and future challenges toward seamless and unified joint compression are discussed. The joint compression framework, which bridges the gap between visual analysis and signal-level representation, is expected to contribute to a series of applications, such as video surveillance and autonomous driving.

\section{Conclusion}
\label{end}
We introduce the progress in VCM standard and the block-based hybrid video coding schemes designed for specific computer vision tasks. Video coding is a set of strategies for spatial domain and temporal domain. When we design video coding schemes for specific computer vision tasks and propose VCM standards, we should also think about temporal domain.


%

\ifCLASSOPTIONcaptionsoff
  \newpage
\fi



%

\bibliographystyle{IEEEtran}
\bibliography{Task_Oriented_Video_Coding_A_Survey}



%








\end{document}